\documentclass[notitlepage,a4paper,aps,prd,onecolumn,superscriptaddress,nofootinbib,longbibliography]{revtex4-1}
\usepackage{amsmath}
\usepackage{amsfonts}
\usepackage{amssymb}
\usepackage[latin1,utf8]{inputenc}
\usepackage[T1]{fontenc}
\usepackage[colorlinks=true]{hyperref}

\usepackage{soul}
\setstcolor{red}

\newcommand{\tp}[1]{\overset{\bullet}{#1}\vphantom{#1}}
\newcommand{\lc}[1]{\overset{\circ}{#1}\vphantom{#1}}

\newcommand{\D}{\mathcal{D}}
\newcommand{\DD}{\mathrm{D}}
\newcommand{\dd}{\mathrm{d}}

\newcommand{\iprod}[1]{\iota_{#1}}

\begin{document}

\title{Scalar-torsion theories of gravity II: $L(T, X, Y, \phi)$ theory}

\author{Manuel Hohmann}
\email{manuel.hohmann@ut.ee}
\affiliation{Laboratory of Theoretical Physics, Institute of Physics, University of Tartu, W. Ostwaldi 1, 50411 Tartu, Estonia}

\author{Christian Pfeifer}
\email{christian.pfeifer@ut.ee}
\affiliation{Laboratory of Theoretical Physics, Institute of Physics, University of Tartu, W. Ostwaldi 1, 50411 Tartu, Estonia}

\begin{abstract}
We consider Lorentz invariant scalar-tensor teleparallel gravity theories with a Lagrangian built from the torsion scalar, a scalar field, its kinetic term and a derivative coupling between the torsion and the scalar field. The field equations of the theory are derived and the relation between the spin connection and the antisymmetric part of the tetrad field equations is found explicitly, which is an important consistency result for Lorentz invariant teleparallel theories of gravity. Afterwards we study the behaviour of this class of theories under conformal transformations and find that such transformations map different theories in this class onto each other.
\end{abstract}

\maketitle

\section{Introduction}\label{sec:intro}
The observation of accelerating phases in the early and late universe~\cite{Ade:2015xua,Ade:2015rim,Ade:2015lrj}, which is not explained by pure general relativity (GR) without the assumption of either a cosmological constant or additional matter besides the directly observed visible matter, has led to the development of a plethora of modified gravity theories aimed at explaining these observations~\cite{Capozziello:2010zz,Clifton:2011jh,Nojiri:2017ncd}.

A particular interesting class of such theories, which has been studied in the recent years, are the so called modified teleparallel theories of gravity. In the teleparallel approach gravity is attributed to torsion, with the fundamental fields given by a tetrad and a flat spin connection, instead of to curvature, with a metric as the fundamental field~\cite{Moller:1961,Aldrovandi:2013wha,Golovnev:2018red}. Nowadays it is well known that general relativity can equally well be formulated in both frameworks~\cite{Maluf:2013gaa}. Modifications of general relativity formulated in the either or the other approach however differ in their mathematical structure as well as in their phenomenological predictions. One prominent modification is the non-minimal coupling of a scalar field either to the curvature~\cite{Fujii:2003pa,Faraoni:2004pi} or to the torsion~\cite{Geng:2011aj,Izumi:2013dca,Chakrabarti:2017moe,Otalora:2013tba,Jamil:2012vb,Chen:2014qsa}.

The question arises whether different instances of scalar-curvature, known as scalar-tensor theories of gravity, or scalar-torsion theories can be related to each other by performing a scalar field dependent conformal transformation of the metric respectively the tetrad, possibly combined with a redefinition of the scalar field. This question has been studied for different classes of scalar-curvature theories~\cite{Flanagan:2004bz,Bettoni:2015wta,Zumalacarregui:2013pma}, while for scalar-torsion theories only few studies exist~\cite{Bamba:2013jqa,Wright:2016ayu}.

This article is the second out of a series of three articles which aims to answer this question. In the first article~\cite{Hohmann:2018vle} we discussed scalar-torsion theories of gravity, where the sole restriction we imposed was that the theory is invariant under local Lorentz transformations and that there is no direct coupling between matter fields and  the spin connection. Apart from that an arbitrary coupling of a scalar field to all other dynamical variables was considered. In particular the behaviour of this most general class of scalar-torsion theories under the aforementioned conformal transformations was investigated. We found that conformal transformations of the tetrad and corresponding redefinitions of the scalar field map scalar-torsion theories onto each other. For more restricted classes of theories, for example those which are specifically based on the torsion scalar used in the teleparallel equivalent of general relativity (TEGR), there exists an obstruction on such transformations to map the theories of the chosen class onto each other. The reason for this difficulty is that under conformal transformations additional kinetic coupling terms appear~\cite{Bamba:2013jqa,Wright:2016ayu}. This can be circumvented by using a different, conformally invariant contraction of the torsion tensor, in analogy to Weyl gravity~\cite{Maluf:2011kf}, or by including the kinetic coupling terms into the action. In this article we choose the latter approach.

In the present paper we restrict the general considerations of the previous paper to a newly introduced and more narrow class of theories, whose gravitational Lagrangian \(L(T, X, Y, \phi)\) is a free function of four scalar terms: the torsion scalar $T$, a scalar field $\phi$, its kinetic term $X$ and a derivative coupling between the torsion and the scalar field $Y$. Henceforth we call such theories $L(T,X,Y,\phi)$-theories. This class of theories has a number of relevant and interesting subclasses and examples, which includes teleparallel dark energy~\cite{Geng:2011aj,Izumi:2013dca,Chakrabarti:2017moe,Otalora:2013tba,Jamil:2012vb,Chen:2014qsa}, scalar-torsion gravity without derivative coupling~\cite{Hohmann:2018rwf} or conformally coupled scalar-torsion gravity~\cite{Bamba:2013jqa}. A specifically interesting subclass, whose action is very similar to that of scalar-tensor theory, will be discussed in the third and final part of this series~\cite{Hohmann:2018ijr}. Our work makes use of the covariant formulation of scalar-torsion formulation of scalar-torsion gravity~\cite{Hohmann:2018rwf}, which is based on the covariant formulation of teleparallel gravity in terms of a tetrad and a flat spin connection, which has to be varied separately~\cite{Krssak:2015oua,Golovnev:2017dox}.

After the more technical study of general scalar torsion theories in our previous work~\cite{Hohmann:2018vle} we now apply the techniques developed there to the aforementioned \(L(T, X, Y, \phi)\) subclass of scalar-tensor theories. In particular we will derive the field equations of the theory, study their relation and obtain the important consistency result for Lorentz invariant teleparallel theories of gravity explicitly. We show that the anti-symmetric part of the tetrad field equation is equal to the equations of motion of the spin connection. This feature was shown for various modified and extended theories of teleparallel gravity~\cite{Golovnev:2017dox,Hohmann:2017duq,Hohmann:2018rwf} and keeps to hold in scalar torsion theories.

As already mentioned, it is an important feature of the previously studied class of scalar-torsion theories that conformal transformations of the tetrad relate different theories to each other, which are contained in the same generic class~\cite{Hohmann:2018vle}. We study the transformation behavior of the scalar quantities $T$, $X$, $Y$ and $\phi$ to show that conformal transformations map similarly different $L(T,X,Y,\phi)$-theories onto each other, so that also this class is closed under these transformations. In particular this means that a conformal transformation does not introduce new terms in the Lagrangian which can not be expressed in terms of these scalars derived from the transformed tetrad.

The outline of this article is as follows. In section~\ref{sec:action} we define all notions and tensors needed to write down the gravitational action we study throughout this article. Afterwards in section~\ref{sec:variation} we discuss how the variation of the action can be performed, before we display the equations of motion for the fundamental fields in section~\ref{sec:feqs}. In section~\ref{sec:conformal} we study the behavior of the theory under conformal transformations. We display examples found in the literature, which belong to the class of theories we discuss, in section~\ref{sec:examples}, before we finally conclude in section~\ref{sec:conclusion}. We supplement our work with two appendices, in which we derive the field equations of the studied class of gravity theories: using differential forms in appendix~\ref{app:conffeqdf}, as well as using components in appendix~\ref{app:conffeqco}.

\section{Dynamical fields and action}\label{sec:action}
We start our discussion of the \(L(T, X, Y, \phi)\) class of scalar-torsion gravity theories by defining their dynamical fields and the general form of the action functional. As usual in teleparallel gravity theories in their covariant formulation the dynamical variables are given by a coframe field \(\theta^a = \theta^a{}_{\mu}\dd x^{\mu}\) and a flat spin connection \(\tp{\omega}^a{}_b = \tp{\omega}^a{}_{b\mu}\dd x^{\mu}\); in addition, we also consider a scalar field \(\phi\). The frame field dual to the coframe field \(\theta^a\) will be denoted \(e_a = e_a{}^{\mu}\partial_{\mu}\). Note that the coframe field induces a metric
\begin{equation}\label{eqn:metric}
g_{\mu\nu} = \eta_{ab}\theta^a{}_{\mu}\theta^b{}_{\nu}\,,
\end{equation}
and we use the sign convention \(\eta_{ab} = \mathrm{diag}(-1,1,1,1)\) for the Minkowski metric. We denote quantities related to the flat spin connection with a bullet (\(\bullet\)), while those related to the Levi-Civita connection \(\lc{\nabla}_{\mu}\) will be denoted with an open circle (\(\circ\)). This in particular applies to the torsion tensor of the flat spin connection
\begin{equation}
T^{\rho}{}_{\mu\nu} = e_a{}^{\rho}\left(\partial_{\mu}e^a{}_{\nu} - \partial_{\nu}e^a{}_{\mu} + \tp{\omega}^a{}_{b\mu}e^b{}_{\nu} - \tp{\omega}^a{}_{b\nu}e^b{}_{\mu}\right)\,.
\end{equation}
The action we consider throughout this article is composed from four scalar functions of the dynamical fields. We start by defining the torsion scalar, which is given by
\begin{equation}\label{eqn:torsscal}
T = \frac{1}{2}T^{\rho}{}_{\mu\nu}S_{\rho}{}^{\mu\nu}\,,
\end{equation}
where we also introduced the superpotential
\begin{equation}\label{eqn:suppot}
S_{\rho\mu\nu} = \frac{1}{2}\left(T_{\nu\mu\rho} + T_{\rho\mu\nu} - T_{\mu\nu\rho}\right) - g_{\rho\mu}T^{\sigma}{}_{\sigma\nu} + g_{\rho\nu}T^{\sigma}{}_{\sigma\mu}
\end{equation}
which is a linear function of the torsion.

We further define the kinetic term of the scalar field as
\begin{equation}\label{eqn:defx}
X = -\frac{1}{2}g^{\mu\nu}\phi_{,\mu}\phi_{,\nu}\,,
\end{equation}
as well as the derivative coupling term
\begin{equation}\label{eqn:defy}
Y = g^{\mu\nu}T^{\rho}{}_{\rho\mu}\phi_{,\nu}\,.
\end{equation}
Finally, the fourth scalar quantity will be the scalar field \(\phi\) itself. The gravitational part of the action is then defined through the most general Lagrangian depending on these four scalar quantities,
\begin{equation}\label{eqn:confaction}
S_g\left[\theta^a, \tp{\omega}^a{}_b, \phi\right] = \int_ML\left(T, X, Y, \phi\right)\theta\dd^4x\,,
\end{equation}
where the volume form is given by
\begin{equation}\label{eqn:vol}
\theta\dd^4x \equiv \det(\theta^a{}_{\mu})\dd^4x = \mathrm{vol}_{\theta} = \theta^0 \wedge \theta^1 \wedge \theta^2 \wedge \theta^3\,.
\end{equation}
Note that the action~\eqref{eqn:confaction} is invariant under combined local Lorentz transformations of the tetrad and the spin connection,
\begin{equation}
\theta^a{}_{\mu} \mapsto \Lambda^a{}_b\theta^b{}_{\mu}\,, \quad
\tp{\omega}^a{}_{b\mu} \to \Lambda^a{}_c\Lambda_b{}^d\tp{\omega}^c{}_{d\mu} - \Lambda_b{}^c\partial_{\mu}\Lambda^a{}_c\,.
\end{equation}
We further consider a matter action of the form
\begin{equation}\label{eqn:mataction}
S_m[\theta^a, \phi, \chi^I]\,,
\end{equation}
which we also assume to be locally Lorentz invariant. It thus in particular follows that the action we use in this article belongs to the class of actions we studied in our previous work~\cite{Hohmann:2018vle}.

\section{Variation}\label{sec:variation}
In order to derive the field equations, we will need the variation of the gravitational and matter action. We start with the matter action, whose variation we now write in the form
\begin{equation}\label{eqn:cmatactvar}
\delta S_m[\theta^a, \phi, \chi^I] = \int_M\left(\Theta_a{}^{\mu}\delta\theta^a{}_{\mu} + \vartheta\delta\phi + \varpi_I\delta\chi^I\right)\theta\dd^4x\,.
\end{equation}
The variation of the action~\eqref{eqn:confaction} takes the form
\begin{equation}\label{eqn:cgravactvar}
\delta S_g\left[\theta^a, \tp{\omega}^a{}_b, \phi\right] = \int_M\left(e_a{}^{\mu}\delta\theta^a{}_{\mu}L + L_T\delta T + L_X\delta X + L_Y\delta Y + L_{\phi}\delta\phi\right)\theta\dd^4x\,,
\end{equation}
where subscripts denote derivatives with respect to the corresponding argument of \(L\). In order to derive the field equations, we need to calculate the variations of the scalar quantities which appear in the action~\eqref{eqn:confaction} with respect to the dynamical fields. We begin with the torsion scalar, whose variation can most conveniently be written in the form
\begin{subequations}
\begin{align}
\delta_{\theta}T &= -2S^{\rho\sigma\mu}T_{\rho\sigma\nu}e_a{}^{\nu}\delta\theta^a{}_{\mu} - 2S_{\rho}{}^{\mu\nu}e_a{}^{\rho}\tp{\D}_{\nu}\delta\theta^a{}_{\mu}\,,\\
\delta_{\omega}T &= \left(T^{\mu}{}_{\rho\sigma}e_a{}^{\rho}e_b{}^{\sigma} - 2T^{\rho}{}_{\rho\nu}e_a{}^{\mu}e_b{}^{\nu}\right)\delta\tp{\omega}^{ab}{}_{\mu}\,,
\end{align}
\end{subequations}
where we introduced the Fock-Ivanenko derivative 
\begin{equation}
\tp{\D}_{\nu}\delta\theta^a{}_{\mu} = \partial_{\nu}\delta\theta^a{}_{\mu} + \tp{\omega}^a{}_{b\nu}\delta\theta^b{}_{\mu}\,.
\end{equation}
For the scalar field kinetic term we find the variation
\begin{subequations}
\begin{align}
\delta_{\phi}X &= -g^{\mu\nu}\phi_{,\nu}\delta\phi_{,\mu}\,,\\
\delta_{\theta}X &= g^{\mu\nu}\phi_{,\nu}\phi_{,\rho}e_a{}^{\rho}\delta\theta^a{}_{\mu}\,,
\end{align}
\end{subequations}
while the variation of the derivative coupling term yields
\begin{subequations}
\begin{align}
\delta_{\phi}Y &= g^{\mu\nu}T^{\rho}{}_{\rho\nu}\delta\phi_{,\mu}\,,\\
\delta_{\theta}Y &= -e_a{}^{\nu}\left(g^{\rho\sigma}T^{\mu}{}_{\nu\rho}\phi_{,\sigma} + 2g^{\mu\rho}T^{\sigma}{}_{\sigma(\nu}\phi_{,\rho)}\right)\delta\theta^a{}_{\mu} + 2g^{\rho[\mu}e_a{}^{\nu]}\phi_{,\rho}\tp{\D}_{\nu}\delta\theta^a{}_{\mu}\,,\\
\delta_{\omega}Y &= e_a{}^{\mu}e_b{}^{\nu}\phi_{,\nu}\delta\tp{\omega}^{ab}{}_{\mu}\,.
\end{align}
\end{subequations}
With these variations we can now derive the field equations. We defer their derivation to the appendix, and directly proceed with showing the results.

\section{Field equations}\label{sec:feqs}
Using the action detailed in the preceding section, we are now in the position to derive the corresponding field equations of the class of theories we consider in this article. Since the derivation of these field equations is rather lengthy, we defer it to the appendix; a version using differential forms and our formerly obtained results~\cite{Hohmann:2018vle} is shown in appendix~\ref{app:conffeqdf}, while a derivation in components is presented in appendix~\ref{app:conffeqco}. Both calculations yield the same result, as one would expect.

From the variation of the action with respect to the coframe field \(\theta^a{}_{\mu}\) we obtain the tetrad field equation
\begin{multline}\label{eqn:confeqtet}
-Le_a{}^{\mu} - 2\lc{\nabla}_{\nu}\left(L_TS_{\rho}{}^{\mu\nu}\right)e_a{}^{\rho} + 2S^{\rho\sigma\mu}T_{\rho\sigma\nu}e_a{}^{\nu}L_T + 2K^b{}_{a\nu}L_TS_{\rho}{}^{\mu\nu}e_b{}^{\rho} - L_Xg^{\mu\nu}\phi_{,\nu}\phi_{,\rho}e_a{}^{\rho}\\
+ 2\lc{\nabla}_{\nu}\left(L_Y\phi_{,\rho}\right)g^{\rho[\mu}e_a{}^{\nu]} + L_Ye_a{}^{\nu}\left(g^{\rho\sigma}T^{\mu}{}_{\nu\rho}\phi_{,\sigma} - 2g^{\mu\rho}T^{\sigma}{}_{\sigma(\nu}\phi_{,\rho)}\right) + 2K^b{}_{a\nu}L_Y\phi_{,\rho}g^{\rho[\mu}e_b{}^{\nu]} = \Theta_a{}^{\mu}\,.
\end{multline}
Alternatively, one can also lower the upper free index, and transform the free Lorentz index into a spacetime index by using the coframe field. This yields the equation
\begin{multline}\label{eqn:confeqtet2}
-Lg_{\mu\nu} - 2\lc{\nabla}_{\rho}\left(L_TS_{\nu\mu}{}^{\rho}\right) - L_T\left(T^{\rho}{}_{\rho\sigma}T^{\sigma}{}_{\mu\nu} + 2T^{\rho}{}_{\rho\sigma}T_{(\mu\nu)}{}^{\sigma} - \frac{1}{2}T_{\mu\rho\sigma}T_{\nu}{}^{\rho\sigma} + T_{\mu\rho\sigma}T^{\rho\sigma}{}_{\nu}\right) - L_X\phi_{,\mu}\phi_{,\nu}\\
+ \lc{\nabla}_{\nu}\left(L_Y\phi_{,\mu}\right) - \lc{\nabla}_{\sigma}\left(L_Y\phi_{,\rho}\right)g^{\rho\sigma}g_{\mu\nu} + L_Y\left(T_{(\mu\nu)}{}^{\rho}\phi_{,\rho} + \frac{1}{2}T^{\rho}{}_{\mu\nu}\phi_{,\rho} + T^{\rho}{}_{\rho\mu}\phi_{,\nu}\right) = \Theta_{\mu\nu}\,,
\end{multline}
where we have also used the symmetry \(\Theta_{[\mu\nu]} = 0\) of the energy-momentum tensor. It is convenient to split these equations into their symmetric and antisymmetric parts. The symmetric part is then given by
\begin{multline}\label{eqn:confeqtets}
-Lg_{\mu\nu} - 2\lc{\nabla}_{\rho}\left(L_TS_{(\mu\nu)}{}^{\rho}\right) + L_TS_{(\mu}{}^{\rho\sigma}T_{\nu)\rho\sigma} - L_X\phi_{,\mu}\phi_{,\nu}\\
+ \lc{\nabla}_{(\mu}\left(L_Y\phi_{,\nu)}\right) - \lc{\nabla}_{\sigma}\left(L_Y\phi_{,\rho}\right)g^{\rho\sigma}g_{\mu\nu} + L_Y\left(T_{(\mu\nu)}{}^{\rho}\phi_{,\rho} + T^{\rho}{}_{\rho(\mu}\phi_{,\nu)}\right) = \Theta_{\mu\nu}\,,
\end{multline}
whereas the antisymmetric part reads
\begin{equation}\label{eqn:confeqcon}
3\partial_{[\rho}L_TT^{\rho}{}_{\mu\nu]} + \partial_{[\mu}L_Y\phi_{,\nu]} - \frac{3}{2}L_YT^{\rho}{}_{[\mu\nu}\phi_{,\rho]} = 0\,.
\end{equation}
Note that the latter agrees with the field equations derived by a restricted variation with respect to the spin connection, which takes into account the flatness of the spin connection. Finally, variation with respect to the scalar field yields the equation
\begin{equation}\label{eqn:confeqscal}
g^{\mu\nu}\lc{\nabla}_{\mu}\left(L_YT^{\rho}{}_{\rho\nu} - L_X\phi_{,\nu}\right) - L_{\phi} = \vartheta\,,
\end{equation}
which completes the set of gravitational field equations. We remark that these are complemented by a set of matter field equations, which read \(\varpi_I = 0\).

\section{Conformal transformations}\label{sec:conformal}
Using the action and field equations derived in the previous sections, we now come to the discussion of conformal transformations. The aim of this section is to show that the class of scalar-torsion theories we consider here retains its form under a conformal transformation of the tetrad and a redefinition of the scalar field of the form
\begin{equation}\label{eqn:conftrans}
\bar{\theta}^a{}_{\mu} = e^{\gamma(\phi)}\theta^a{}_{\mu}\,, \quad \bar{e}_a{}^{\mu} = e^{-\gamma(\phi)}e_a{}^{\mu}\,, \quad \bar{\phi} = f(\phi)\,.
\end{equation}
For this purpose we need to calculate the transformation behavior of the four scalar quantities introduced in section~\ref{sec:action}, which can be derived from the transformation of the torsion, contortion and superpotential tensors, which read
\begin{equation}\label{eqn:tkstrans}
\bar{T}^{\rho}{}_{\mu\nu} = T^{\rho}{}_{\mu\nu} - 2\gamma'\delta^{\rho}_{[\mu}\phi_{,\nu]}\,, \quad
\bar{K}_{\mu\nu}{}^{\rho} = K_{\mu\nu}{}^{\rho} - 2\gamma'\delta^{\rho}_{[\mu}\phi_{,\nu]}\,, \quad
\bar{S}_{\rho}{}^{\mu\nu} = e^{-2\gamma}\left(S_{\rho}{}^{\mu\nu} + 4\gamma'\delta^{[\mu}_{\rho}g^{\nu]\sigma}\phi_{,\sigma}\right)\,,
\end{equation}
where we keep in mind that \(f\) and \(\gamma\) are functions of the scalar field \(\phi\). Note that indices on the transformed (barred) tensors must be raised and lowered using the corresponding metric \(\bar{g}_{\mu\nu} = e^{2\gamma}g_{\mu\nu}\). It then follows that the scalars in the action transform as
\begin{equation}\label{eqn:scaltrans}
\bar{T} = e^{-2\gamma}\left(T + 4\gamma'Y + 12(\gamma')^2X\right)\,, \quad
\bar{Y} = e^{-2\gamma}f'(Y + 6\gamma'X)\,, \quad
\bar{X} = e^{-2\gamma}(f')^2X\,, \quad
\bar{\phi} = f\,.
\end{equation}
Remarkably the conformal transformation of the tetrad and the spin connection yields that the scalar quantities $\bar T, \bar X, \bar Y$ and $\bar \phi$ all become functions of the corresponding scalars $T,X,Y$ and $\phi$ of the untransformed tetrad, without any further terms appearing.

With the transformation behaviour obtained we now consider a new, different action functional \(\bar{S}\), which has the same structure as the original action introduced in section~\ref{sec:action}, but with a different gravitational Lagrangian \(\bar{L}\) in its gravitational part \(\bar{S}_g\) and different matter part \(\bar{S}_m\). We then evaluate this new action at the transformed fields,
\begin{equation}
\bar{S}_g\left[\bar{\theta}^a, \tp{\omega}^a{}_b, \bar{\phi}\right] = \int_M\bar{L}\left(\bar{T}, \bar{X}, \bar{Y}, \bar{\phi}\right)\bar{\theta}\dd^4x\,, \quad
\bar{S}_m\left[\bar{\theta}^a, \bar{\phi}, \chi^I\right]\,.
\end{equation}
Substituting the original fields for the transformed fields using the transformation rules~\eqref{eqn:conftrans}, and hence also the rules~\eqref{eqn:scaltrans}, we find that the new action reproduces the original action,
\begin{equation}
\bar{S}_g\left[\bar{\theta}^a, \tp{\omega}^a{}_b, \bar{\phi}\right] = S_g\left[\theta^a, \tp{\omega}^a{}_b, \phi\right]\,, \quad
\bar{S}_m\left[\bar{\theta}^a, \bar{\phi}, \chi^I\right] = S_m\left[\theta^a, \phi, \chi^I\right]\,,
\end{equation}
if and only if their gravitational Lagrangians are related by
\begin{equation}\label{eqn:lagrangetrans}
L\left(T, X, Y, \phi\right) = \bar{L}\left[e^{-2\gamma}\left(T + 4\gamma'Y + 12(\gamma')^2X\right), e^{-2\gamma}f'(Y + 6\gamma'X), e^{-2\gamma}(f')^2X, f\right]\,,
\end{equation}
while the matter actions must satisfy
\begin{equation}\label{eqn:matacttrans}
S_m\left[\theta^a, \phi, \chi^I\right] = \bar{S}_m\left[e^{\gamma(\phi)}\theta^a, f(\phi), \chi^I\right]\,.
\end{equation}
Thus, we see that the action retains is form. It follows that the class of theories we consider in this article is closed under conformal transformations of the tetrad and redefinitions of the scalar field, i.e., by applying a transformation of the form~\eqref{eqn:conftrans} to the field variables and a corresponding transformation of the action as shown above, we obtain another theory belonging to the same class.

Further, note that the variations of the transformed variables are given by
\begin{equation}
\delta\bar{\theta}^a{}_{\mu} = e^{\gamma}\left(\delta\theta^a{}_{\mu} + \gamma'\theta^a{}_{\mu}\delta\phi\right)\,, \quad
\delta\bar{\phi} = f'\delta\phi\,.
\end{equation}
This allows us to write the variation of the matter part \(\bar{S}_m\) of the transformed action in terms of the original variables,
\begin{equation}
\begin{split}
\delta\bar{S}_m\left[\bar{\theta}^a, \bar{\phi}, \chi^I\right] &= \int_M\left(\bar{\Theta}_a{}^{\mu}\delta\bar{\theta}^a{}_{\mu} + \bar{\vartheta}\delta\bar{\phi} + \bar{\varpi}_I\delta\chi^I\right)\bar{\theta}\dd^4x\\
&= \int_M\left[e^{\gamma}\bar{\Theta}_a{}^{\mu}\delta\theta^a{}_{\mu} + \left(\gamma'\bar{\theta}^a{}_{\mu}\bar{\Theta}_a{}^{\mu} + f'\bar{\vartheta}\right)\delta\phi + \bar{\varpi}_I\delta\chi^I\right]e^{4\gamma}\bar{\theta}\dd^4x\,.
\end{split}
\end{equation}
By comparison with the variation~\eqref{eqn:cmatactvar} of the original matter action one thus finds that the matter terms in the field equations transform as
\begin{equation}\label{eqn:matvartrans1}
\Theta_a{}^{\mu} = e^{5\gamma}\bar{\Theta}_a{}^{\mu}\,, \quad
\vartheta = e^{4\gamma}(\gamma'\bar{\theta}^a{}_{\mu}\bar{\Theta}_a{}^{\mu} + f'\bar{\vartheta})\,, \quad
\varpi_I = e^{4\gamma}\bar{\varpi}_I\,.
\end{equation}
Finally, lowering the upper index and transforming the Lorentz index into a spacetime index using the corresponding tetrads on each side of the first equation allows rewriting the transformation of the energy-momentum tensor and its trace as
\begin{equation}\label{eqn:matvartrans2}
\Theta_{\mu\nu} = e^{2\gamma}\bar{\Theta}_{\mu\nu}\,, \quad
\Theta = e^{4\gamma}\bar{\Theta}\,.
\end{equation}
These relations, together with the transformation~\eqref{eqn:lagrangetrans} of the Lagrangian, allow a transformation of the field equations detailed in section~\ref{sec:feqs}. A lengthy, but straightforward calculation shows that they are indeed invariant under this transformation.

This concludes our discussion of conformal transformations, and also of scalar-torsion theories with a single scalar field. Most of the results we obtained can easily be generalized to theories with multiple scalar fields. This will be done in the next section.

\section{Generalization to multiple scalar fields}\label{sec:multi}
In the previous sections we have considered a single scalar field coupled to torsion. It is not too difficult to generalize our considerations to an arbitrary number of scalar fields by the introduction of convenient notations. The numerous terms which then appear in the gravitational Lagrangian can thereby be written in an comprehensive way, as we will discuss in detail in section~\ref{ssec:maction}. The generalized field equations are shown in section~\ref{ssec:mfeqs}. Finally, we discuss conformal transformations and scalar field redefinitions with multiple scalar fields in section~\ref{ssec:mconformal}.

\subsection{Dynamical fields and action}\label{ssec:maction}
Recall that we defined two functions \(X\) and \(Y\) containing derivatives of the scalar field by the relations~\eqref{eqn:defx} and~\eqref{eqn:defy}, which enter the gravitational part of the action. In a theory with multiple fields these terms can be generalized. Introducing the definitions
\begin{equation}
X^{AB} = -\frac{1}{2}g^{\mu\nu}\phi^A_{,\mu}\phi^B_{,\nu}\,, \quad Y^A = T_{\mu}{}^{\mu\nu}\phi^A_{,\nu}\,,
\end{equation}
we can write all possible terms appearing in a compact comprehensive way.
We then allow for the gravitational action~\eqref{eqn:confaction} to depend on all of these terms, all scalar fields and the torsion, such that it takes the form
\begin{equation}\label{eqn:multiconfaction}
S_g\left[\theta^a, \tp{\omega}^a{}_b, \phi^A\right] = \int_ML\left(T, X^{AB}, Y^A, \phi^A\right)\theta\dd^4x\,.
\end{equation}
In addition the matter action is now allowed to depend on all scalar fields. As a result, its variation~\eqref{eqn:cmatactvar} now takes the more general form
\begin{equation}\label{eqn:multicmatactvar}
\delta S_m[\theta^a, \phi^A, \chi^I] = \int_M\left(\Theta_a{}^{\mu}\delta\theta^a{}_{\mu} + \vartheta_A\delta\phi^A + \varpi_I\delta\chi^I\right)\theta\dd^4x\,.
\end{equation}
The variation~\eqref{eqn:cgravactvar} similarly generalizes and takes the form
\begin{equation}\label{eqn:multicgravactvar}
\delta S_g\left[\theta^a, \tp{\omega}^a{}_b, \phi^A\right] = \int_M\left(e_a{}^{\mu}\delta\theta^a{}_{\mu}L + L_T\delta T + L_{X^{AB}}\delta X^{AB} + L_{Y^A}\delta Y^A + L_{\phi^A}\delta\phi^A\right)\theta\dd^4x\,.
\end{equation}
Particular care must be given to the term \(L_{X^{AB}}\delta X^{AB}\), since by definition \(X^{AB}\) is symmetric, \(X^{[AB]} = 0\), and so its components are not independent. Hence, the derivative of \(L\) should not be interpreted as an ordinary derivative with respect to the components of \(X^{AB}\), but as a variational derivative in the sense
\begin{equation}
\delta_XL\left(T, X^{AB}, Y^A, \phi^A\right) = L_{X^{AB}}\left(T, X^{AB}, Y^A, \phi^A\right)\delta X^{AB} = \left.\frac{d}{d\epsilon}L\left(T, X^{AB} + \epsilon\delta X^{AB}, Y^A, \phi^A\right)\right|_{\epsilon = 0}\,.
\end{equation}
Note in particular that \(L_{X^{AB}}\) inherits the symmetry from \(X^{AB}\), such that \(L_{X^{[AB]}} = 0\).

\subsection{Field equations}\label{ssec:mfeqs}
With these definitions in place it is now straightforward to derive the field equations. We omit the derivation here for brevity, and provide the field equations in their final form only, decomposed into symmetric and antisymmetric parts. For the symmetric part~\eqref{eqn:confeqtets} we find
\begin{multline}\label{eqn:multiconfeqtets}
-Lg_{\mu\nu} - 2\lc{\nabla}_{\rho}\left(L_TS_{(\mu\nu)}{}^{\rho}\right) + L_TS_{(\mu}{}^{\rho\sigma}T_{\nu)\rho\sigma} - L_{X^{AB}}\phi^A_{,\mu}\phi^B_{,\nu}\\
+ \lc{\nabla}_{(\mu}\left(L_{Y^A}\phi^A_{,\nu)}\right) - \lc{\nabla}_{\sigma}\left(L_{Y^A}\phi^A_{,\rho}\right)g^{\rho\sigma}g_{\mu\nu} + L_{Y^A}\left(T_{(\mu\nu)}{}^{\rho}\phi^A_{,\rho} + T^{\rho}{}_{\rho(\mu}\phi^A_{,\nu)}\right) = \Theta_{\mu\nu}\,,
\end{multline}
while the antisymmetric part~\eqref{eqn:confeqcon} reads
\begin{equation}\label{eqn:multiconfeqcon}
3\partial_{[\rho}L_TT^{\rho}{}_{\mu\nu]} + \partial_{[\mu}L_{Y^A}\phi^A_{,\nu]} - \frac{3}{2}L_{Y^A}T^{\rho}{}_{[\mu\nu}\phi^A_{,\rho]} = 0\,,
\end{equation}
and the scalar field equations take the form
\begin{equation}\label{eqn:multiconfeqscal}
g^{\mu\nu}\lc{\nabla}_{\mu}\left(L_{Y^A}T^{\rho}{}_{\rho\nu} - L_{X^{AB}}\phi^B_{,\nu}\right) - L_{\phi^A} = \vartheta_A\,.
\end{equation}
Here we have used the aforementioned symmetry of \(L_{X^{AB}}\) in order to suppress symmetrization brackets around the scalar field indices. Due to the introduction of the terms $X^{AB}$ and $Y^A$ the field equations look formally very similar to the equations obtained in the one scalar field case.

\subsection{Conformal transformations}\label{ssec:mconformal}
Finally, we also return to the discussion of invariance of this class of theories under conformal transformations of the tetrad and scalar field redefinitions, which are of the form
\begin{equation}\label{eqn:multiconftrans}
\bar{\theta}^a{}_{\mu} = e^{\gamma(\boldsymbol{\phi})}\theta^a{}_{\mu}\,, \quad \bar{e}_a{}^{\mu} = e^{-\gamma(\boldsymbol{\phi})}e_a{}^{\mu}\,, \quad \bar{\phi}^A = f^A(\boldsymbol{\phi})\,.
\end{equation}
in the case of multiple scalar fields. For this purpose, we need to generalize the transformation behavior~\eqref{eqn:tkstrans} of the torsion, contortion and superpotential tensors, which reads
\begin{equation}\label{eqn:multitkstrans}
\bar{T}^{\rho}{}_{\mu\nu} = T^{\rho}{}_{\mu\nu} - 2\gamma_{,A}\delta^{\rho}_{[\mu}\phi^A_{,\nu]}\,, \quad
\bar{K}_{\mu\nu}{}^{\rho} = K_{\mu\nu}{}^{\rho} - 2\gamma_{,A}\delta^{\rho}_{[\mu}\phi^A_{,\nu]}\,, \quad
\bar{S}_{\rho}{}^{\mu\nu} = e^{-2\gamma}\left(S_{\rho}{}^{\mu\nu} + 4\gamma_{,A}\delta^{[\mu}_{\rho}g^{\nu]\sigma}\phi^A_{,\sigma}\right)\,.
\end{equation}
This further yields the transformations~\eqref{eqn:scaltrans} of the scalar terms in the action, which is now given by
\begin{subequations}
\begin{align}\label{eqn:multiscaltrans}
\bar{T} &= e^{-2\gamma}\left(T + 4\gamma_{,A}Y^A + 12\gamma_{,A}\gamma_{,B}X^{AB}\right)\,,\\
\bar{Y}^A &= e^{-2\gamma}\frac{\partial\bar{\phi}^A}{\partial\phi^B}\left(Y^B + 6\gamma_{,C}X^{BC}\right)\,,\\
\bar{X}^{AB} &= e^{-2\gamma}\frac{\partial\bar{\phi}^A}{\partial\phi^C}\frac{\partial\bar{\phi}^B}{\partial\phi^D}X^{CD}\,,\\
\bar{\phi}^A &= f^A\,.
\end{align}
\end{subequations}
We see that also in this case the four scalar functions which enter the Lagrangian show a closed transformation behavior, in the sense that after the conformal transformation, they can be expressed as functions of each other and no new terms appear. Hence, it follows that under the transformation~\eqref{eqn:multiconftrans} the action~\eqref{eqn:multiconfaction} retains its form, in the same way as for a single scalar field.

As a final remark observe that the transformation of the matter terms in the field equation, which are derived from a variation of the matter action, follow a similar set of rules as in the case of a single scalar field. The only difference from the transformation behavior~\eqref{eqn:matvartrans1} and~\eqref{eqn:matvartrans2} arises from the fact that the scalar field source term \(\vartheta_A\) now carries a scalar field index, and so the transformation law becomes
\begin{equation}
\vartheta_A = e^{4\gamma}\left(\gamma_{,A}\bar{\theta}^a{}_{\mu}\bar{\Theta}_a{}^{\mu} + \frac{\partial\bar{\phi}^B}{\partial\phi^A}\bar{\vartheta}_B\right)\,,
\end{equation}
while the remaining transformation laws change only implicitly due to the fact that the conformal transformation parameter \(\gamma\) now depends on all scalar fields.

This concludes our general discussion of (multi-)scalar-torsion theories. In the next section we will connect our results to a number of specific examples and subclasses of theories studied elsewhere in the literature.

\section{Examples}\label{sec:examples}
After discussing general aspects of the class of \(L(T,X,Y,\phi)\) gravity theories discussed in this article, we now connect our results to a few classes of gravity theories which are being discussed in the literature, and have peculiar properties. In section~\ref{ssec:conf} we consider the case in which the scalar field is conformally coupled. Further, in section~\ref{ssec:stg} we consider a class of theories which is constructed in analogy to scalar-tensor gravity, so that both have many aspects in common. Finally, in section~\ref{ssec:noder} we discuss a recently studied class of theories without derivative couplings.

\subsection{Conformally coupled scalar field}\label{ssec:conf}
In analogy to theories based on the curvature of the Levi-Civita connection, one may discuss a class of theories in which the scalar field is conformally coupled~\cite{Bamba:2013jqa}. The action of these theories can be written in the form
\begin{equation}
S_g = \frac{1}{2\kappa^2}\int_M\left[f(T) + \frac{1}{2}g^{\mu\nu}\phi_{,\mu}\phi_{,\nu} - \frac{C}{2}\phi^2T - D\phi T_{\mu}{}^{\mu\nu}\phi_{,\nu} - \frac{\phi^{m + 1}}{m + 1}\right]\theta\dd^4x\,,
\end{equation}
and their cosmology has been studied in the literature. The peculiar property of this class of models is the fact that they are invariant under a simultaneous conformal transformation of the tetrad and rescaling of the scalar field.

\subsection{Scalar-torsion analogue of scalar-tensor gravity}\label{ssec:stg}
A particularly interesting subclass of the class of gravity theories we studied in this article is given by a Lagrangian \(L(T,X,Y,\phi)\) that is linear in its first three arguments, while keeping an arbitrary dependence on the scalar field \(\phi\). The action can then be brought to the form
\begin{equation}\label{eqn:classaction}
S = \frac{1}{2\kappa^2}\int_M\left[-\mathcal{A}(\phi)T + 2\mathcal{B}(\phi)X + 2\mathcal{C}(\phi)Y - 2\kappa^2\mathcal{V}(\phi)\right]\theta\dd^4x + S_m\left[e^{\alpha(\phi)}\theta^a, \chi^I\right]\,,
\end{equation}
with free functions \(\mathcal{A}, \mathcal{B}, \mathcal{C}, \mathcal{V}\) of the scalar field defining the gravitational part of the action. For the matter part of the action we consider a coupling of the matter fields to a conformally related tetrad, which is defined through another free function \(\alpha\). One can show that this action retains its form under conformal transformations, so that one can study a number of different conformal frames; however, this would exceed the scope of this article, and so we defer the full investigation of these aspects to another work~\cite{Hohmann:2018ijr}.

\subsection{Scalar-torsion theory without derivative couplings}\label{ssec:noder}
Another interesting subclass is obtained if one allows for an arbitrary coupling between the torsion scalar and the scalar field, but restricts the Lagrangian to be linear in \(X\) and independent of \(Y\), hence disallowing derivative couplings. If one further disallows a direct coupling between the scalar and matter fields, one obtains an action of the form
\begin{equation}\label{eqn:ndactiong}
S = \frac{1}{2\kappa^2}\int_M\left[F(T,\phi) - 2Z(\phi)X\right]\theta\dd^4x + S_m[\theta^a, \chi^I]\,.
\end{equation}
A peculiar aspect of this class of theory is the fact that the connection field equations~\eqref{eqn:confeqcon} take a simple form, which is similar to that of \(f(T)\) gravity, and which allows for a number of generic solutions for the spin connection, which is independent of the functions \(f\) and \(Z\); see~\cite{Hohmann:2018rwf} for a detailed discussion.

\section{Conclusion}\label{sec:conclusion}
In this second article on scalar-torsion theories we applied the general formalism developed in \cite{Hohmann:2018vle} to the class of theories which is built from the torsion scalar, a scalar field, its kinetic term and a derivative coupling between the torsion and the scalar field. This class includes interesting models studied in the literature, as we discussed in section \ref{sec:examples}, and extends the analysis of such models to the most general possible Lagrangian which can be written down with the variables mentioned.

For this class of extended theories of gravity we derived the field equations presented in section \ref{sec:feqs}, which we derived by two methods: in differential form language as well as in coordinate components. The field equations \eqref{eqn:confeqtets}, \eqref{eqn:confeqcon} and~\eqref{eqn:confeqscal} demonstrate explicitly, what was proven on general abstract ground in the previous paper of this series, namely how the variation of the action with respect to the tetrads is connected to the variation of the action with respect to the spin connection. The spin connection equation of motion is the anti-symmetric part of the tetrad equation of motion and only those pairs of spin connection and tetrads which solve both field equations can be considered as solution of the theory.

An important feature of gravity theories including a non-minimal coupled scalar field is their behaviour under conformal transformations of the metric, here the tetrad, with a function depending on the scalar field. We found the transformation behaviour of the constituents of the gravity action in \ref{sec:conformal} and remarkably they can be expressed as function  of the same set of scalars obtained from the conformally transformed tetrad. This means starting from a gravity theory defined by a Lagrangian depending on the torsion scalar, a scalar field, its kinetic term and a derivative coupling between the torsion and the scalar field, the theory is still of this type in terms of the transformed fields after a conformal transformation. However the Lagrangian as function of the fields changes throughout this process and so conformal transformations relate different scalar-torsion theories to each other.

All results presented in this article were derived in detail for scalar-torsion theories containing one scalar field. With the introduction of a suitable counting index which allows to define convenient shorthand notations in section \ref{sec:multi} we could easily generalize all obtained results to multiple scalar fields coupled to each other and to the torsion. All of our findings from the single scalar field carry over to the multiple scalar field considerations.

In summary this paper demonstrates explicitly how the theoretical framework developed in \cite{Hohmann:2018vle} builds the ground for the analysis of Lorentz and diffeomorphism invariant scalar-torsion theories, which makes these theories available as viable alternatives to scalar-tensor theories formulated in terms of the metric.

On the conceptual side an interesting question for the future is to derive the Hamilton formulation of scalar-torsion theories to determine their propagating degrees of freedom and their initial value formulation. If this is possible in the generality in which the theories are discussed here is an open question. For the examples mentioned this is certainly doable.

From the phenomenological point of view, the next steps are to derive observables from this class of theories to find viability constraints on the Lagrangian in consideration. On cosmological scales dark energy phenomenology is one source of relevant observables. Particular classes of theories discussed here will be addressed in the context of cosmology  in the next article of this series \cite{Hohmann:2018ijr}. Further constraints every modified or extended theory of gravity has to pass are the PPN constraints obtained in the solar system.

Moreover it will be most exciting to derive the propagation of gravitational waves in this class of theories and to compare their propagation speed with the speed of light, whose equal speed was recently confirmed.

\begin{acknowledgments}
The authors thank Martin Krššák for helpful comments and discussions. They gratefully acknowledge the full financial support of the Estonian Ministry for Education and Science through the Institutional Research Support Project IUT02-27 and Startup Research Grant PUT790, as well as the European Regional Development Fund through the Center of Excellence TK133 ``The Dark Side of the Universe''.
\end{acknowledgments}

\appendix

\section{Derivation of the $L(T, X, Y, \phi)$ field equations using differential forms}\label{app:conffeqdf}
Here we show how to derive the field equations of the theory in terms of differential forms, as it was done in the general consideration of scalar-torsion actions in \cite{Hohmann:2018vle}. Recall the field equations in differential form language
\begin{align}
	\tp{\DD}\Pi_a - \Upsilon_a = \Sigma_a\label{eqn:tetradformeq}\\
	\tp{\DD}\Pi^{[a}\wedge \theta^{b]} - \Upsilon^{[a}\wedge \theta^{b]} = 0\label{eqn:spinconnformeq}\\
	\Phi = - \Psi\,.\label{eqn:scalarformeq}
\end{align}
The $3$-forms $\Sigma_a$ and the $4$-form $\Psi$ are obtained from the variation of the matter action.

The further differential forms appearing can be obtained by writing  the variation of the gravity action~\eqref{eqn:confaction} as
\begin{align}\label{eqn:varform}
	\delta S_g = \int \Upsilon_a \wedge \delta \theta^a + \Pi_a \wedge \delta T^a + \Phi \delta \phi\,,
\end{align}
where $\Upsilon_a$ are $3$-forms, $\Pi_a$ are $2$-forms and $\Phi$ is a $4$-form. To identify these differential forms observe that in a first step the variation of action is
\begin{align}
	\delta S_g = \int_M \bigg( L_{T} \delta T + L_X \delta X + L_Y \delta Y + L_\phi \delta \phi \bigg) \mathrm{vol}_{\theta} + L \delta \mathrm{vol}_{\theta}\,,
\end{align}
where subscripts on $L$ denote partial derivative of $L$ with respect to the subscript. Further expansion of the variation can be done by using the equations \eqref{eqn:torsscal}, \eqref{eqn:defx}, \eqref{eqn:defy} and \eqref{eqn:vol} which define $T,X,Y,\phi$ and $\mathrm{vol}_{\theta}$. Moreover it is useful to express the torsion scalar in terms of the local and linear gravitational constitutive relation (or supermetric)~\cite{Ong:2013qja,Ferraro:2016wht,Itin:2016nxk,Hohmann:2017duq}
\begin{align}
T = \frac{1}{4}T^{a}{}_{cd}T^{b}{}_{ef}\chi_{a}{}^{cd}{}_{b}{}^{ef}
\end{align}
with
\begin{align}\label{eqn:TEGRconst}
\chi_{a}{}^{cd}{}_{b}{}^{ef}
=2 \Bigg( \delta_{b}^{[d}\eta^{c][e} \delta_a^{f]} + \frac{1}{2}\eta_{ab}\eta^{c[e}\eta^{{f}]d} + 2 \delta_a^{[c} \eta^{d][e} \delta_{b}^{f]}\Bigg)\,.
\end{align}
Denoting the interior product between vectorfields $X$ and differential forms $\Omega$ as $\iprod{X}\Omega$ we find
\begin{align}
	\delta T^a
	&= \frac{1}{2}\delta T^{a}{}_{cd}T^{b}{}_{ef}\chi_{a}{}^{cd}{}_{b}{}^{ef} = S_a{}^{cd}(\iprod{e_d}\iprod{e_c}\delta T^a + 2 T^a{}_{e[c}\iprod{e_{d]}}\delta\theta^e)\\
	\delta X
	&= - \eta^{ab}\ \iprod{e_a}\dd\phi\ ( \ \iprod{e_b}\dd\delta\phi\  - \ \iprod{e_c}\dd\phi\ \iprod{e_b}\delta \theta^c )\\
	\delta Y
	&= \eta^{ab}(\ \iprod{e_a}\dd\phi\ [\iprod{e_b}\iprod{e_c}\delta T^c + 2 T^c{}_{e[c}\iprod{e_{b]}}\delta \theta^e] + T^c{}_{cb}[ \ \iprod{e_a}\dd\delta\phi\  - \ \ \iprod{e_d}\dd\phi\ \ \iprod{e_a}\delta\theta^d ] )\\
	\delta \mathrm{vol}_{\theta}
	&= \delta \theta^a \wedge \iprod{e_a}\mathrm{vol}_\theta\,.
\end{align}
Collecting all terms the variation of the action becomes
\begin{align}
	&\int_M \bigg(L_T S_a{}^{cd} + L_Y \eta^{fd} \ \iprod{e_f}\dd\phi\  \delta^c_a\bigg) \iprod{e_d}\iprod{e_c}\delta T^a\ \mathrm{vol}_\theta +\bigg( \eta^{ab}(L_Y T^c{}_{ca} - L_X \ \iprod{e_a}\dd\phi\ ) \bigg) \iprod{e_b}\dd \delta\phi\ \mathrm{vol}_{\theta}+ L\ \delta\theta^a \wedge \iprod{e_a}\mathrm{vol}_{\theta} \nonumber\\
	&+ \bigg(2 L_T S_a{}^{cd} T^a{}_{bc} + L_X \eta^{ad} \iprod{e_a}\dd\phi\ \iprod{e_b}\dd\phi\  - L_Y \Big[ \eta^{da}T^c{}_{ca} \ \iprod{e_b}\dd\phi\  - \eta^{ad}\ \iprod{e_a}\dd\phi\  T^c{}_{bc} + \eta^{ac} \ \iprod{e_a}\dd\phi\  T^d{}_{bc} \Big] \bigg) \iprod{e_d}\delta\theta^b\ \mathrm{vol}_{\theta}\,.
\end{align}
To bring the variation of the action  into the form \eqref{eqn:varform} we apply the product rule for the interior product $\iprod{X}(\alpha \wedge \beta) = \iprod{X}\alpha \wedge \beta + (-1)^p \alpha \wedge \iprod{X}\beta$ for $\alpha$ being a $p$-form, to the differential forms in the above equation and use integration by parts to find
\begin{align}
		&\int_M -\bigg(L_T S_a{}^{cd} + L_Y \eta^{fd} \ \ \iprod{e_f}\dd\phi\ \  \delta^c_a\bigg) \iprod{e_d}\iprod{e_c}\mathrm{vol}_\theta \wedge \delta T^a
		- \dd\bigg( \eta^{ab}(L_Y T^c{}_{ca} - L_X \ \iprod{e_a}\dd\phi\ ) \iprod{e_b}\mathrm{vol}_{\theta}\bigg)  \delta\phi\nonumber\\
		&- \bigg(2 L_T S_a{}^{cd} T^a{}_{bc} + L_X \eta^{ad} \ \iprod{e_a}\dd\phi\ \ \iprod{e_b}\dd\phi\  - L_Y \Big[ \eta^{da}T^c{}_{ca} \ \iprod{e_b}\dd\phi\  - \eta^{ad}\ \iprod{e_a}\dd\phi\  T^c{}_{bc} + \eta^{ac} \ \iprod{e_a}\dd\phi\  T^d{}_{bc} \Big] - L\bigg) \iprod{e_d} \mathrm{vol}_{\theta} \wedge \delta\theta^b\,.
\end{align}
In this form we can easily read off the desired differential forms
\begin{align}
	\Upsilon_a
	&= - \bigg(2 L_T S_a{}^{cd} T^a{}_{bc} + L_X \eta^{ad} \ \iprod{e_a}\dd\phi\ \ \iprod{e_b}\dd\phi\  - L_Y \Big[ \eta^{da}T^c{}_{ca} \ \iprod{e_b}\dd\phi\  - \eta^{ad}\ \iprod{e_a}\dd\phi\  T^c{}_{bc} + \eta^{ac} \ \iprod{e_a}\dd\phi\  T^d{}_{bc} \Big] - L\bigg) \iprod{e_d} \mathrm{vol}_{\theta}\,,\\
	\Pi_a
	&=-\bigg(L_T S_a{}^{cd} + L_Y \eta^{fd} \ \ \iprod{e_f}\dd\phi\ \  \delta^c_a\bigg) \iprod{e_d}\iprod{e_c}\mathrm{vol}_\theta\,,\\
	\Phi
	&=- \lc{\DD}\bigg( \eta^{ab}(L_Y T^c{}_{ca} - L_X \ \iprod{e_a}\dd\phi\ ) \iprod{e_b}\mathrm{vol}_{\theta}\bigg) = - \iprod{e_b}\lc{\DD}\bigg( \eta^{ab}(L_Y T^c{}_{ca} - L_X \ \iprod{e_a}\dd\phi\ ) \bigg)\mathrm{vol}_{\theta}\,,
\end{align}
which define the field equations \eqref{eqn:tetradformeq}, \eqref{eqn:spinconnformeq} and \eqref{eqn:scalarformeq}. In the last line we used that
\begin{align}
	\dd (Q^b\ \iprod{e_b}\mathrm{vol}_{\theta}) = \lc{\DD}(Q^b\ \iprod{e_b}\mathrm{vol}_{\theta}) = \lc{\DD}Q^b \wedge \iprod{e_b}\mathrm{vol}_{\theta} =  \iprod{e_b}\lc{\DD}Q^b\ \mathrm{vol}_{\theta}\,.
\end{align}
The first equality is due to the fact that $Q^b\ \iprod{e_b}\mathrm{vol}_{\theta}$ has no free Lorentz indices, the second uses that $\lc{\DD}(\iprod{e_b}\mathrm{vol}_{\theta})$ vanishes since the Levi-Civita covariant derivative has no torsion and the last one uses the product rule for the interior product again.

Do evaluate the tetrad field equation \eqref{eqn:tetradformeq} observe that with help of the contortion $K^a{}_b$ we can write $\tp{\DD}\Pi_a = \lc{\DD}\Pi_a - K^q{}_a \wedge \Pi_q$ and that $\lc{\DD}(\iprod{e_b}\iprod{e_a}\mathrm{vol}_{\theta})=0$, due to the vanishing torsion of the Levi-Civita connections, as already mentioned above.

\section{Derivation of the $L(T, X, Y, \phi)$ field equations using components}\label{app:conffeqco}
In this appendix we show how the field equations discussed in section~\ref{sec:feqs} can be derived using components with respect to a coordinate basis. We split the derivation into several steps. We derive the tetrad field equation~\eqref{eqn:confeqtet} and its equivalent form~\eqref{eqn:confeqtet2} in section~\ref{sapp:cotet}, the connection field equation~\eqref{eqn:confeqcon} in section~\ref{sapp:cocon} and the scalar field equation~\eqref{eqn:confeqscal} in section~\ref{sapp:coscal}. Finally, in section~\ref{sapp:corel} we decompose the tetrad field equations into their symmetric and antisymmetric parts, and show that the antisymmetric part reproduces the connection field equations.

\subsection{Tetrad field equations}\label{sapp:cotet}
We start with the derivation of the tetrad field equations~\eqref{eqn:confeqtet}. For this purpose we write the variation of the action~\eqref{eqn:confaction} in the form
\begin{equation}
\begin{split}
\delta_eS &= \int_M\left(e_a{}^{\mu}\delta\theta^a{}_{\mu}L + L_T\delta_eT + L_X\delta_eX + L_Y\delta_eY + \Theta_a{}^{\mu}\delta\theta^a{}_{\mu}\right)\theta\dd^4x\\
&= \int_M\bigg\{\bigg[Le_a{}^{\mu} - 2S^{\rho\sigma\mu}T_{\rho\sigma\nu}e_a{}^{\nu}L_T - L_Ye_a{}^{\nu}\left(g^{\rho\sigma}T^{\mu}{}_{\nu\rho}\phi_{,\sigma} + 2g^{\mu\rho}T^{\sigma}{}_{\sigma(\nu}\phi_{,\rho)}\right)\\
&\phantom{=}+ L_Xg^{\mu\nu}\phi_{,\nu}\phi_{,\rho}e_a{}^{\rho} + \Theta_a{}^{\mu}\bigg]\delta\theta^a{}_{\mu} - 2\left[L_TS_{\rho}{}^{\mu\nu}e_a{}^{\rho} - L_Yg^{\rho[\mu}e_a{}^{\nu]}\phi_{,\rho}\right]\tp{\D}_{\nu}\delta\theta^a{}_{\mu}\bigg\}\theta\dd^4x\,.
\end{split}
\end{equation}
In order to eliminate the Fock-Ivanenko derivative on the variation of the tetrad, we make use of the integration by parts formula
\begin{equation}
\int_MV_a{}^{[\mu\nu]}\tp{\D}_{\nu}\delta\theta^a{}_{\mu}\theta\dd^4x = \int_M\left(-\lc{\nabla}_{\nu}V_{\rho}{}^{[\mu\nu]} + K^{\sigma}{}_{\rho\nu}V_{\sigma}{}^{[\mu\nu]}\right)e_a{}^{\rho}\delta\theta^a{}_{\mu}\theta\dd^4x\,.
\end{equation}
Integration by parts thus yields the variation
\begin{multline}
\delta_eS = \int_M\bigg[Le_a{}^{\mu} - 2S^{\rho\sigma\mu}T_{\rho\sigma\nu}e_a{}^{\nu}L_T - L_Ye_a{}^{\nu}\left(g^{\rho\sigma}T^{\mu}{}_{\nu\rho}\phi_{,\sigma} + 2g^{\mu\rho}T^{\sigma}{}_{\sigma(\nu}\phi_{,\rho)}\right) + L_Xg^{\mu\nu}\phi_{,\nu}\phi_{,\rho}e_a{}^{\rho}\\
+ 2\lc{\nabla}_{\nu}\left(L_TS_{\rho}{}^{\mu\nu}\right)e_a{}^{\rho} - 2K^b{}_{a\nu}L_TS_{\rho}{}^{\mu\nu}e_b{}^{\rho} - 2\lc{\nabla}_{\nu}\left(L_Y\phi_{,\rho}\right)g^{\rho[\mu}e_a{}^{\nu]} + 2K^b{}_{a\nu}L_Y\phi_{,\rho}g^{\rho[\mu}e_b{}^{\nu]} + \Theta_a{}^{\mu}\bigg]\delta\theta^a{}_{\mu}\theta\dd^4x\,,
\end{multline}
which immediately yields the field equations~\eqref{eqn:confeqtet}. By lowering the upper index and transforming the Lorentz index into a spacetime index using the coframe field, one obtains the equivalent field equations~\eqref{eqn:confeqtet2}.

\subsection{Connection field equations}\label{sapp:cocon}
In order to derive the connection field equations~\eqref{eqn:confeqcon}, it is most convenient to make use of the relation
\begin{equation}
T = -\lc{R} + 2\lc{\nabla}_{\nu}T_{\mu}{}^{\mu\nu}
\end{equation}
between the Ricci scalar of the Levi-Civita connection and the teleparallel torsion scalar. Since the Ricci scalar and the connection coefficients of the Levi-Civita connection are independent of the teleparallel spin connection \(\tp{\omega}^a{}_{b\mu}\), we find that the variation can be written as
\begin{equation}
\delta_{\omega}T = 2\lc{\nabla}_{\nu}\delta_{\omega}T_{\mu}{}^{\mu\nu} = 2\lc{\nabla}_{\nu}\left(e_a{}^{\mu}e_b{}^{\nu}\delta\tp{\omega}^{ab}{}_{\mu}\right)\,.
\end{equation}
Further, note that the only allowed variations of the spin connection are of the form \(\delta\tp{\omega}^a{}_{b\mu} = \tp{\D}_{\mu}\pi^a{}_b\) with \(\pi_{(ab)} = 0\), in order to preserve the flatness of the spin connection. Thus, we can write the variation of the action as
\begin{equation}
\begin{split}
\delta_{\omega}S &= \int_M\left(L_T\delta_{\omega}T + L_Y\delta_{\omega}Y\right)\theta\dd^4x\\
&= \int_M\left[2L_T\lc{\nabla}_{\nu}\left(e_a{}^{\mu}e_b{}^{\nu}\tp{\D}_{\mu}\pi^{ab}\right) + L_Ye_a{}^{\mu}e_b{}^{\nu}\phi_{,\nu}\tp{\D}_{\mu}\pi^{ab}\right]\theta\dd^4x\\
&= -\int_M\left(2\partial_{\nu}L_T - L_Y\phi_{,\nu}\right)e_a{}^{\mu}e_b{}^{\nu}\tp{\D}_{\mu}\pi^{ab}\theta\dd^4x\,.
\end{split}
\end{equation}
We now rewrite the partial derivative on \(L_T\) as a teleparallel covariant derivative, and make use of the integration by parts formula
\begin{equation}
\int_MV^{\mu}{}_{ab}\tp{\D}_{\mu}\pi^{[ab]}\theta\dd^4x = \int_M\left[-\tp{\nabla}_{\mu}V^{\mu}{}_{\rho\sigma} + K^{\mu}{}_{\nu\mu}V^{\nu}{}_{\rho\sigma}\right]e_a{}^{\rho}e_b{}^{\sigma}\pi^{[ab]}\theta\dd^4x\,,
\end{equation}
in order to eliminate the Fock-Ivanenko derivative on the variation term. This yields the variation
\begin{equation}
\delta_{\omega}S = \int_M\left[\tp{\nabla}_{\mu}\left(2\tp{\nabla}_{\nu}L_T - L_Y\tp{\nabla}_{\nu}\phi\right) - K^{\rho}{}_{\mu\rho}\left(2\tp{\nabla}_{\nu}L_T - L_Y\tp{\nabla}_{\nu}\phi\right)\right]e_a{}^{\mu}e_b{}^{\nu}\pi^{ab}\theta\dd^4x\,.
\end{equation}
Keeping in mind that \(\pi^{ab}\) is antisymmetric, we find that the connection field equation is the antisymmetric part of the expression in square brackets. It can further be simplified by using the relation
\begin{equation}
\tp{\nabla}_{\mu}\tp{\nabla}_{\nu}\psi - \tp{\nabla}_{\nu}\tp{\nabla}_{\mu}\psi = T^{\rho}{}_{\nu\mu}\partial_{\rho}\psi
\end{equation}
for the commutator of teleparallel covariant derivatives on a scalar function. The field equations can then finally be written as
\begin{equation}
\begin{split}
0 &= \tp{\nabla}_{[\mu}L_Y\tp{\nabla}_{\nu]}\phi + L_Y\tp{\nabla}_{[\mu}\tp{\nabla}_{\nu]}\phi - 2\tp{\nabla}_{[\mu}\tp{\nabla}_{\nu]}L_T + L_Y\tp{\nabla}_{[\mu}\phi K^{\rho}{}_{\nu]\rho} - 2\tp{\nabla}_{[\mu}L_TK^{\rho}{}_{\nu]\rho}\\
&= \partial_{[\mu}L_Y\phi_{,\nu]} - \frac{1}{2}L_Y\left(T^{\rho}{}_{\mu\nu}\phi_{,\rho} + 2T^{\rho}{}_{\rho[\mu}\phi_{,\nu]}\right) + T^{\rho}{}_{\mu\nu}\partial_{\rho}L_T + 2T^{\rho}{}_{\rho[\mu}\partial_{\nu]}L_T\\
&= 3\partial_{[\rho}L_TT^{\rho}{}_{\mu\nu]} + \partial_{[\mu}L_Y\phi_{,\nu]} - \frac{3}{2}L_YT^{\rho}{}_{[\mu\nu}\phi_{,\rho]}\,.
\end{split}
\end{equation}
These are the field equations~\eqref{eqn:confeqcon}.

\subsection{Scalar field equation}\label{sapp:coscal}
Variation with respect to the scalar field yields
\begin{equation}
\begin{split}
\delta_{\phi}S &= \int_M\left(L_X\delta_{\phi}X + L_Y\delta_{\phi}Y + L_{\phi}\delta\phi + \vartheta\delta\phi\right)\theta\dd^4x\\
&= \int_M\left[g^{\mu\nu}\left(L_YT^{\rho}{}_{\rho\nu} - L_X\phi_{,\nu}\right)\delta\phi_{,\mu} + (L_{\phi} + \vartheta)\delta\phi\right]\theta\dd^4x\\
&= \int_M\left[g^{\mu\nu}\lc{\nabla}_{\mu}\left(L_X\phi_{,\nu} - L_YT^{\rho}{}_{\rho\nu}\right) + L_{\phi} + \vartheta\right]\delta\phi\theta\dd^4x\,,
\end{split}
\end{equation}
where we have applied integration by parts to obtain the last line. The scalar field equation thus reads
\begin{equation}\label{eqn:scalarcomp}
g^{\mu\nu}\lc{\nabla}_{\mu}\left(L_X\phi_{,\nu} - L_YT^{\rho}{}_{\rho\nu}\right) + L_{\phi} + \vartheta = 0\,.
\end{equation}
We see that we obtain the scalar field equation~\eqref{eqn:confeqscal}.

\subsection{Relation between field equations}\label{sapp:corel}
We finally decompose the tetrad field equations~\eqref{eqn:confeqtet2} into their symmetric and antisymmetric parts. The symmetric part~\eqref{eqn:confeqtets} is straightforward to derive by making use of the relation
\begin{equation}
S_{(\mu}{}^{\rho\sigma}T_{\nu)\rho\sigma} = -2T^{\rho}{}_{\rho\sigma}T_{(\mu\nu)}{}^{\sigma} + \frac{1}{2}T_{\mu\rho\sigma}T_{\nu}{}^{\rho\sigma} - T^{\rho\sigma}{}_{(\mu}T_{\nu)\rho\sigma}\,.
\end{equation}
Therefore, we focus on the antisymmetric part, which reads
\begin{equation}
\begin{split}
0 &= 2\lc{\nabla}_{\rho}\left(L_TS_{[\nu\mu]}{}^{\rho}\right) + L_T\left(T^{\rho}{}_{\rho\sigma}T^{\sigma}{}_{\mu\nu} + T^{\rho\sigma}{}_{[\nu}T_{\mu]\rho\sigma}\right) - \lc{\nabla}_{[\nu}\left(L_Y\phi_{,\mu]}\right) - L_Y\left(\frac{1}{2}T^{\rho}{}_{\mu\nu}\phi_{,\rho} + T^{\rho}{}_{\rho[\mu}\phi_{,\nu]}\right)\\
&= 2\partial_{\rho}L_TS_{[\nu\mu]}{}^{\rho} + L_T\left(3\lc{\nabla}_{[\rho}T^{\rho}{}_{\mu\nu]} + T^{\rho}{}_{\rho\sigma}T^{\sigma}{}_{\mu\nu} - T^{\rho\sigma}{}_{[\mu}T_{\nu]\rho\sigma}\right) + \partial_{[\mu}L_Y\phi_{,\nu]} - \frac{3}{2}L_YT^{\rho}{}_{[\mu\nu}\phi_{,\rho]}\,.
\end{split}
\end{equation}
For the first term we find
\begin{equation}
2\partial_{\rho}L_TS_{[\nu\mu]}{}^{\rho} = 3\partial_{[\rho}L_TT^{\rho}{}_{\mu\nu]}\,.
\end{equation}
Further, we calculate
\begin{equation}
\begin{split}
\lc{\nabla}_{[\rho}T^{\rho}{}_{\mu\nu]} &= \partial_{[\rho}T^{\rho}{}_{\mu\nu]} + \lc{\Gamma}^{\rho}{}_{\sigma[\rho}T^{\sigma}{}_{\mu\nu]} - \lc{\Gamma}^{\sigma}{}_{[\mu\rho}T^{\rho}{}_{|\sigma|\nu]} - \lc{\Gamma}^{\sigma}{}_{[\nu\rho}T^{\rho}{}_{\mu]\sigma}\\
&= \partial_{[\rho}T^{\rho}{}_{\mu\nu]} + \tp{\Gamma}^{\rho}{}_{\sigma[\rho}T^{\sigma}{}_{\mu\nu]} - K^{\rho}{}_{\sigma[\rho}T^{\sigma}{}_{\mu\nu]}\\
&= \frac{2}{3}\tp{R}^{\rho}{}_{[\mu\nu]\rho} - \frac{1}{3}\left(T^{\rho}{}_{\mu\nu}T^{\sigma}{}_{\sigma\rho} - T^{\rho\sigma}{}_{[\mu}T_{\nu]\rho\sigma}\right)\\
&= -\frac{1}{3}\left(T^{\rho}{}_{\mu\nu}T^{\sigma}{}_{\sigma\rho} - T^{\rho\sigma}{}_{[\mu}T_{\nu]\rho\sigma}\right)\,,
\end{split}
\end{equation}
where we have used the facts that the Levi-Civita connection has vanishing torsion and the teleparallel connection has vanishing curvature, which can be expressed as
\begin{equation}
\lc{\Gamma}^{\rho}{}_{[\mu\nu]} = 0\,, \quad \tp{R}^{\sigma}{}_{\mu\nu\rho} = 0\,.
\end{equation}
Hence, we obtain
\begin{equation}
3\lc{\nabla}_{[\rho}T^{\rho}{}_{\mu\nu]} + T^{\rho}{}_{\mu\nu}T^{\sigma}{}_{\sigma\rho} - T^{\rho\sigma}{}_{[\mu}T_{\nu]\rho\sigma} = 0\,,
\end{equation}
so that the antisymmetric part of the field equations finally reduces to
\begin{equation}
0 = 3\partial_{[\rho}L_TT^{\rho}{}_{\mu\nu]} + \partial_{[\mu}L_Y\phi_{,\nu]} - \frac{3}{2}L_YT^{\rho}{}_{[\mu\nu}\phi_{,\rho]}\,.
\end{equation}
This is simply the connection field equation~\eqref{eqn:confeqcon}.

\bibliography{scaltors}
\end{document}